\documentclass[a4paper,aps,prb,reprint,showpacs]{revtex4-1}
\usepackage[latin9]{inputenc}
\usepackage{color}
\usepackage{amsmath}
\usepackage{amssymb}
\usepackage{graphicx}
\usepackage{esint}

\makeatletter

\pdfpageheight\paperheight
\pdfpagewidth\paperwidth

\usepackage{mathrsfs}
\usepackage{enumitem}
\usepackage{tabu}

\newcommand{\ket}[1]{\left|#1\right\rangle}

\newcommand{\ii}{\text{i}}
\newcommand{\dd}{\text{d}}

\allowdisplaybreaks[4]

\newcommand{\thickhline}{%
    \noalign {\ifnum 0=`}\fi \hrule height 1.25pt
    \futurelet \reserved@a \@xhline
}
\newcolumntype{"}{@{\hskip\tabcolsep\vrule width 1.25pt\hskip\tabcolsep}}

\makeatother

\begin{document}

\title{Topological order in the Kitaev/Majorana chain in the presence of disorder and interactions }

\author{Niklas M. Gergs}
\affiliation{Institute for Theoretical Physics, Center for Extreme Matter and
Emergent Phenomena, Utrecht University, Leuvenlaan 4, 3584 CE Utrecht,
The Netherlands}
\author{Lars Fritz}
\affiliation{Institute for Theoretical Physics, Center for Extreme Matter and
Emergent Phenomena, Utrecht University, Leuvenlaan 4, 3584 CE Utrecht,
The Netherlands}
\author{Dirk Schuricht}
\affiliation{Institute for Theoretical Physics, Center for Extreme Matter and
Emergent Phenomena, Utrecht University, Leuvenlaan 4, 3584 CE Utrecht,
The Netherlands}

\date{\today}
\begin{abstract}
We study the combined effect of interactions and disorder on topological order in one dimension. To this end we consider a generalized Kitaev chain including fermion-fermion interactions and disorder in the chemical potential. We determine the phase diagram by performing density-matrix renormalization group calculations on the corresponding spin-1/2 chain. We find that moderate disorder or repulsive interactions individually stabilize the topological order, which remains valid for their combined effect. However, both repulsive and attractive interactions lead to a suppression of the topological phase at strong disorder. 
\end{abstract}


\pagestyle{plain}
\maketitle

\section{Introduction}\label{sec:intro} 
Partially motivated by their possible applications in quantum information technology, recent years have seen a huge interest\cite{Wilczek09,Alicea12,LeijnseFlensberg12,Beenakker13,ElliotFranz15} in the realization of Majorana fermions in condensed-matter systems both from theoretical and experimental groups. One recent experimental work~\cite{Nadj-Perge-14} performed scanning tunneling experiments on ferromagnetic atomic chains on a superconducting substrate and observed a strongly localized conductance signal around the edges indicative of Majorana edge modes. In fact, the localization at the edges was even stronger than expected.\cite{Dumitrescu-15,Peng-15} In this context the interactions in the system played an important role, i.e., the experimental system was non-surprisingly more complicated than the simple Kitaev chain\cite{Kitaev01} usually serving as a toy model for the emergence of Majorana edge modes in solid-state systems. 

There has been a broad interest in how experimental deviations from the clean, non-interacting setup affect the Majorana edge states in the Kitaev chain as well as systems akin to it. Two of the most relevant experimental influences are disorder\cite{Brouwer-11,Lobos-12,DeGottardi-13,DeGottardi-13prb,Pientka-13,Altland-14,Crepin-14,Xue-15} and interactions.\cite{Gangadharaiah-11,Stoudenmire-11,Sela-11,ChengTu11,LutchynFisher11,Lobos-12,HS12,Thomale-13,Manolescu-14,Crepin-14,Rahmani-15a,Kells15,Milsted-15,Chan-15,KST15,Kells15b,Rahmani-15b,OBrienWright15} Concerning the effects of disorder it has been observed that moderate disorder supports the topological phase by pinning down quasiparticles associated to the phase transition, which has especially been investigated for the two-dimensionsional toric code.\cite{Tsomokos-11,WottonPachos11,BravyiKoenig12} Similarly, it has been shown\cite{Stoudenmire-11,Sela-11,HS12,Thomale-13,Manolescu-14} that repulsive interactions generically broaden the window of chemical potentials over which the topological phase and thus Majorana fermions exist. Much less is known on the combined effect of disorder and interactions, which has so far only been investigated in topological quantum wires. Performing a renormalization-group analysis of the corresponding bosonized low-energy theory in replica space, Lobos et al.~\cite{Lobos-12} showed that disorder and repulsive interactions reinforce each other in suppressing the topological phase and thus eliminating Majorana edge modes in the wires. Cr\'{e}pin et al.\cite{Crepin-14} corroborated this finding using a Gaussian variational approach.

Here we will analyze the combined effect of disorder and interactions in a microscopic model, namely a generalized Kitaev/Majorana chain. Specifically, we consider a one-dimensional chain of spinless fermions in the presence of p-wave superconducting pairing, nearest-neighbor interactions, and a disordered chemical potential. While the special cases of the interacting model  without disorder\cite{Sela-11,HS12,Thomale-13,KST15} as well as the non-interacting, disordered\cite{DeGottardi-13,DeGottardi-13prb,Altland-14} model have been studied previously and can be investigated by analytical methods, the combined effect of interactions and disorder is not amenable to analytic methods. Thus we employ the density matrix renormalization group (DMRG) method\cite{Schollwoeck11} to calculate several observables from which we determine the phase boundary between the topological and trivial phases. We find that moderate disorder stabilizes the topological order in the non-interacting as well as interacting model. However, strong disorder leads to a suppression of the topological phase which is amplified by both repulsive and attractive interactions.

This article is organized as follows: In the next section we define the system and discuss its main properties, including the appearance of Majorana edge states and its mapping to a spin chain. In Sec.~\ref{sub:Interacting-case-without-disorder} we present results on the interacting system without disorder, before we recall known results on the non-interacting model with disorder in Sec.~\ref{sec:Non-interacting-case-with}. In Sec.~\ref{sec:Possible-criteria} we discuss possible criteria for the detection of the topological phase from our numerical DMRG simulations, the details of which we present in Sec.~\ref{sub:Implementation}. In Sec.~\ref{sec:Interplay-of-disorder} we present our main results on the phase diagram in the presence of interactions and disorder, see Figs.~\ref{fig:disorder-interaction-compare} and~\ref{fig:pd2}, before we conclude.

\section{System}
In order to investigate the interplay of disorder and interactions and their influence on the stability of topological phases we consider an extension of the well-known Kitaev/Majorana chain~\cite{Kitaev01}, namely
\begin{eqnarray}
H&=&-\sum_{i=1}^{N-1}\left(t\,c_{i}^{\dagger}c_{i+1}-\Delta\,c_{i}c_{i+1}+\text{h.c.}\right)\nonumber\\
&&+U\sum_{i=1}^{N-1}\bigl(2\,c_{i}^{\dagger}c_{i}-1\bigr)\bigl(2\,c_{i+1}^{\dagger}c_{i+1}-1\bigr)\nonumber\\
&&-\sum_{i=1}^{N}\mu_{i}\left(c_{i}^{\dagger}c_{i}-\frac{1}{2}\right),\label{eq:Kitaev-Hamiltonian}
\end{eqnarray}
where the operators $c_i$ and $c_i^\dagger$ annihilate or create a spinless fermion at lattice site $i$ respectively. The first term describes hopping of the fermions between neighboring sites, the second is a p-wave superconducting pairing, and the third term represents an interaction term between fermions on neighboring sites. The corresponding amplitudes $t$, $\Delta$ and $U$ are assumed to be real (furthermore we assume $t$ and $\Delta$ to be positive) and constant along the chain. In contrast, the chemical potential $\mu_i$ in the last term is allowed to be site dependent and thus to contain disorder. Throughout this manuscript we assume the values $\mu_i$ to stem from a uniform disorder distribution with mean value $\bar{\mu}$ and variance $\sigma_\mu^2$, i.e., the $\mu_i$ are uniformly distributed in the interval $\bar{\mu}-\sqrt{3}\sigma_\mu<\mu_i<\bar{\mu}+\sqrt{3}\sigma_\mu$. The generalization to other disorder distributions is straightforward; we do not expect any qualitative changes of our results. Unless stated differently, we assume an open chain of length $N$.

For the later discussion it is useful to rewrite the Hamiltonian \eqref{eq:Kitaev-Hamiltonian} in two different ways (see e.g. Ref.~\onlinecite{Fendley12}). First, we can introduce two Majorana fermions $\gamma_{i,a}$ and $\gamma_{i,b}$ at each lattice site via $c_i=(\gamma_{i,a}+\ii\gamma_{i,b})/2$. The Majorana fermions satisfy the algebra $\gamma_{i,a}^\dagger=\gamma_{i,a}$, $\gamma_{i,b}^\dagger=\gamma_{i,b}$, $\{\gamma_{i,a},\gamma_{j,a}\}=\{\gamma_{i,b},\gamma_{j,b}\}=2\delta_{ij}$, $\{\gamma_{i,a},\gamma_{j,b}\}=0$ and $\gamma_{i,a}^2=\gamma_{i,b}^2=1$. In terms of these the Hamiltonian reads 
\begin{eqnarray}
H&=&\frac{\ii}{2}\sum_{i=1}^{N-1}\left[(t+\Delta)\gamma_{i,b}\gamma_{i+1,a}-(t-\Delta)\gamma_{i,a}\gamma_{i+1,b}\right]\nonumber\\
&&-U\sum_{i=1}^{N-1}\gamma_{i,a}\gamma_{i,b}\gamma_{i+1,a}\gamma_{i+1,b}-\frac{\ii}{2}\sum_{i=1}^{N}\mu_{i}\gamma_{i,a}\gamma_{i,b}.\label{eq:Majorana-Hamiltonian}
\end{eqnarray}
The appearance of Majorana edge zero modes is most easily seen in the homogeneous, non-interacting model at $t=\Delta$. Denoting the chemical potential by $\mu$ it is straightforward to show\cite{Fendley12} that the operator 
\begin{equation}
\Psi_\text{left}=\sum_{i=1}^N\left(-\frac{\mu}{2t}\right)^{i-1}\gamma_{i,a}
\end{equation}
commutes with the Hamiltonian up to terms exponentially small in the system size provided $|\mu|<2t$. Obviously, $\Psi_\text{left}$ is localized at the left boundary; the Majorana edge mode at the right boundary can be constructed analogously. The condition $|\mu|<2t$ is also sufficient\cite{DeGottardi-13prb} for the existence of Majorana edge modes for $\Delta\neq t$ (as long as $\Delta\neq 0$) and hence determines the topological phase in the clean, non-interacting system. 

Alternatively, the model \eqref{eq:Kitaev-Hamiltonian} can be mapped to a spin-chain Hamiltonian via the Jordan--Wigner transformation $\sigma_i^x=\prod_{j<i}(1-2c_j^\dagger c_j)(c_i^\dagger+c_i)$, $\sigma_i^y=-\ii\prod_{j<i}(1-2c_j^\dagger c_j)(c_i^\dagger-c_i)$, $\sigma_i^z=2c_i^\dagger c_i-1=\ii\gamma_{i,a}\gamma_{i,b}$ with the result
\begin{eqnarray}
H&=&-\sum_{i=1}^{N-1}\left[\frac{t+\Delta}{2}\sigma_i^x\sigma_{i+1}^x+\frac{t-\Delta}{2}\sigma_i^y\sigma_{i+1}^y-U\sigma_i^z\sigma_{i+1}^z\right]\nonumber\\
&&-\frac{1}{2}\sum_{i=1}^{N}\mu_{i}\sigma_i^z,
\label{eq:Spin-Hamiltonian}
\end{eqnarray}
where the $\sigma_i^a$, $a=x,y,z$, denote the usual Pauli matrices. In the homogeneous, non-interacting limit this simplifies to the well-known XY model in a magnetic field. For $|\mu|<2t$ this model is in its ordered phase with $\langle\sigma_i^x\rangle\neq 0$, which corresponds to the topological phase of the Kitaev chain. The relation between the local order parameter in the spin-chain representation \eqref{eq:Spin-Hamiltonian} and the non-local topological order in the Kitaev chain \eqref{eq:Majorana-Hamiltonian} is given by the non-local Jordan--Wigner transformation.\cite{Fendley12}

In the following sections we discuss the phase diagram of the clean, interacting system followed by the non-interacting, disordered one. We then turn to the investigation of their combined effect. 

\section{Interacting case without disorder}\label{sub:Interacting-case-without-disorder}
The phase diagram of the interacting Kitaev chain without disorder, i.e., the model \eqref{eq:Kitaev-Hamiltonian} with $\mu_i=\mu$, has been  determined\cite{Sela-11,HS12,Thomale-13,KST15} employing the equivalent spin-chain representation \eqref{eq:Spin-Hamiltonian}. It was shown that the topological and trivial phases of the non-interacting model extend to weak and moderate repulsive and attractive interaction strengths. At large repulsions, $U>t$, additional incommensurate and commensurate charge-density wave phases (Mott insulator) exist. 

In this article we focus on weak and moderate interactions. In order to obtain quantitative results on the phase boundary between the topological and trivial phase we diagonalize the non-interacting Hamiltonian (assuming periodic boundary conditions) using a standard Bogoliubov transformation (see e.g. Ref.~\onlinecite{Calabrese-12jsm1}). The critical chemical potential $\mu_\text{c}$ is then determined in perturbation theory in $U$ from the condition that the excitation gap closes; the result reads\cite{footnote1}
\begin{eqnarray}
\mu_{\text{c}} &=& 2t-\frac{8Ut}{\pi(\Delta^2-t^2)^{3/2}}\biggl[\sqrt{\Delta^2-t^2}\,\Delta\nonumber\\
& &\qquad-(2\Delta^2-t^2)\ln\frac{\Delta+\sqrt{\Delta^2-t^2}}{t}\biggr]+\mathcal{O}(U^2)\nonumber\\
&\approx&2t+\left[\frac{32}{3\pi}-\frac{32}{15\pi}\frac{\Delta-t}{t}\right]U+\mathcal{O}(U^2),
\label{eq:linear-extrpolated-delta}
\end{eqnarray}
where in the last line we have expanded also to leading order in $(\Delta-t)/t$. The perturbative result \eqref{eq:linear-extrpolated-delta} has been confirmed numerically. The linear approximation with respect to $(\Delta-t)/t$ works surprisingly well even for appreciable deviations of the order of $|\Delta-t|\sim t/2$. We stress that repulsive interactions stabilize the topological phase in the sense that $\mu_\text{c}$ increases with $U$. 

We note that a pair coupling deviating from $\Delta=t$ causes no shift of $\mu_{\text{c}}$ in the absence of interactions while for the interacting model the sign of the shift also depends on the sign of the interaction $U$. The absence of any effect of $\Delta\neq t$ on $\mu_\text{c}$ in the non-interacting case can be understood from the Majorana representation \eqref{eq:Majorana-Hamiltonian}: The coupling $\propto(t-\Delta)$ binds Majorana fermions on neighboring fermionic sites $i$ and $i+1$ and thus acts qualitatively similar to the first term in favoring the formation of Majorana edge modes. Finite interactions, however, influence this binding between neighboring sites, thus causing a correction in $\mu_\text{c}$. 

\section{Non-interacting case with disorder}\label{sec:Non-interacting-case-with}
Now let us turn to the non-interacting model with disorder in the chemical potential. The phase diagram can be obtained by considering a semi-infinite chain and applying a transfer-matrix approach\cite{DeGottardi-13prb,Altland-14} to search for states that decay when moving away from the edge. This leads to the criterion for the existence of Majorana edge states
\begin{equation}
T=\prod_{i=1}^{N}\left(\begin{array}{cc}
-\frac{\mu_{i}}{t+\Delta} & -\frac{t-\Delta}{t+\Delta}\\[2mm]
1 & 0
\end{array}\right), \qquad \max\{|\lambda_{T}|\}<1,
\label{eq:transfermatrix}
\end{equation}
where $\lambda_T$ denotes the eigenvalues of $T$. For the special case $t=\Delta$ the structure of the matrices multiplied in the construction of $T$ simplifies and one obtains in the infinite-size limit
\begin{equation}
\int\dd\mu\,p\left(\mu,\bar{\mu}=\mu_{\text{c}},\sigma_\mu\right)\,\ln\left|\frac{\mu}{2}\right| =\ln t,
\label{eq:non-interacting-infinite}
\end{equation}
where $p\left(\mu,\bar{\mu},\sigma_\mu\right)$ denotes the probability distribution for the $\mu_i$ with mean value $\bar{\mu}$ and variance $\sigma_\mu^2$. For special probability distributions like the uniform, Gaussian, or Cauchy--Lorentz distribution the left-hand side of \eqref{eq:non-interacting-infinite} can be evaluated analytically. In the spin-chain context the condition \eqref{eq:non-interacting-infinite} determines the phase boundary in the quantum Ising chain with disordered transverse field.\cite{ShankarMurthy87,Fisher92,Fisher95} If additionally the tunnel coupling $t$ is disordered (but uncorrelated to the disorder in $\mu_i$), one has to include a disorder average on the right-hand side of \eqref{eq:non-interacting-infinite} as well. 

\begin{figure}[t]
\begin{centering}
\includegraphics[width=0.95\columnwidth]{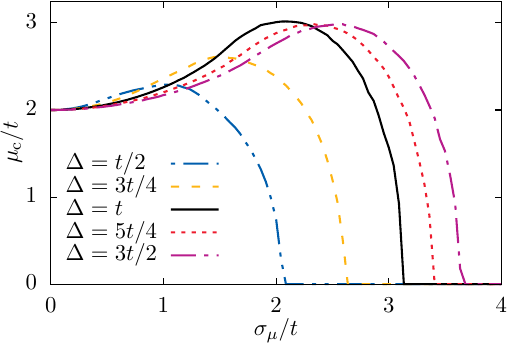}
\par\end{centering}
\caption{(Color online) Phase boundary $\mu_\text{c}$ for different values of the pairing strength $\Delta$. If, for a given standard deviation $\sigma_\mu$, the mean chemical potential $\bar{\mu}$ is below (above) the critical value $\mu_\text{c}$, the system is in the topological (trivial) phase. The data are obtained from the transfer-matrix approach considering $N=10^{6}$ sites.}
\label{fig:non-interacting}
\end{figure}
The criterion on the eigenvalues of the transfer matrix \eqref{eq:transfermatrix} can also be used the determine the phase boundary for $\Delta\neq t$, although no closed result replacing \eqref{eq:non-interacting-infinite} exist. We plot the resulting phase diagram for different values of $\Delta$ in Fig.~\ref{fig:non-interacting}. We observe that under a reduction of $\Delta$ the topological phase shrinks, while increasing $\Delta$ stabilizes it against stronger disorder in the chemical potential. We also observe that the value of $\mu_\text{c}$ at $\sigma_\mu=0$ is unaffected as predicted by the result \eqref{eq:linear-extrpolated-delta}.

\section{Possible criteria}\label{sec:Possible-criteria}
In the remainder of this paper we will analyze the interplay of interactions and disorder in the chemical potential. To this end we will calculate several observables using the density-matrix renormalization group (DMRG) method.\cite{Schollwoeck11} In this section we give a list of  possible criteria to determine the phase boundary between the topological and trivial phase and discuss their practical applicability to the problem at hand. 

\begin{figure}[t]
\begin{center}
\includegraphics[width=0.9\columnwidth]{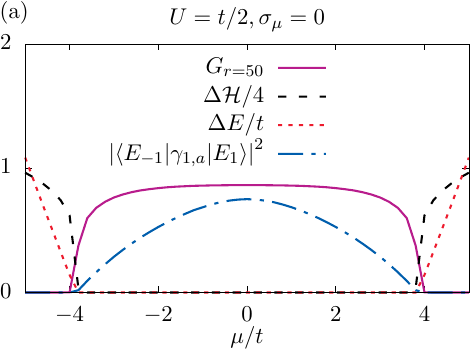}\vspace{5mm}
\includegraphics[width=0.9\columnwidth]{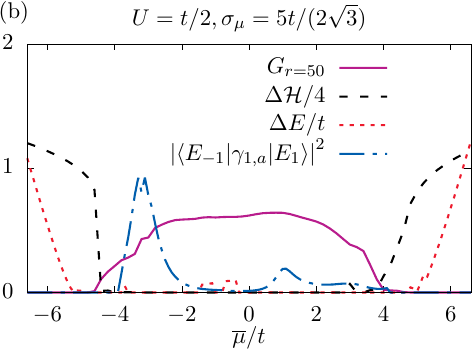}
\end{center}
\caption{(Color online) Function $G_{r=50}$ [see Eq.~\eqref{eq:our-parameter}], entanglement gap $\Delta\mathcal{H}$, energy gap $\Delta E$ above the ground state, and Majorana edge state overlap \eqref{eq:MESO}. (a) Clean system with $\mu_i=\mu$. (b) Exemplary disorder configuration with mean value $\bar{\mu}$. While in the clean system the four criteria agree and give the correct value $\mu=\mu_\text{c}\approx 4t$ for the phase boundary, in the disordered case the energy gap shows significant deviations due to localized zero-energy bulk states while the Majorana edge state overlap possesses strong fluctuations. On the other hand, the correlation function \eqref{eq:our-parameter} and the entanglement gap give the same result. The simulations were performed with $t=\Delta$, $N=200$, and $\chi=32$ kept states.}
\label{fig:mu-dependence}
\end{figure}

1. The simplest possible criterion is the closing of the energy gap between the ground state and the first excited state. This criterion works very well for the clean system as shown in Fig.~\ref{fig:mu-dependence}(a). However, the presence of disorder may lead to localized zero-energy modes, resulting in the closing of the energy gap without a phase transition in the bulk of the system [see Fig.~\ref{fig:mu-dependence}(b)], thus rendering the criterion unreliable. A related criterion, namely the compressibility  $-\partial^{2}E_{\text{GS}}/\partial\mu^{2}$, was discussed recently\cite{Chan-15} and shown to be insensitive to finite-size effects. However, in the disordered case the compressibility shows the same problems as the energy gap. Thus we conclude that both criteria are not suitable for the determination of the phase boundary in the presence of disorder.   

2. The second possible criterion is the relative change of the wave function overlap when changing $\mu$---related to the fidelity\cite{Bauer-10}---$\Delta\psi=\lim_{\varepsilon\searrow0}(1-|\langle\psi_{\mu}|\psi_{\mu+\varepsilon}\rangle|)/\varepsilon$. This parameter also turns out to be very stable in the clean case but gets insensitive to the phase transition for strong disorder as it also tends to be sensitive for the closure of the energy gap. 

3. Another possible criterion is based on the non-interacting picture\cite{Stoudenmire-11} discussed above. One considers the overlap $\left\langle E_{-1}|\gamma_{1,a}|E_{1}\right\rangle$ where $\ket{E_{\pm 1}}$ denotes the  ground state in the even or odd parity sector, respectively, which are connected via the zero-energy Majorana edge state associated with $\gamma_{1,a}$. Hence, this expectation value will be of order one in the topological phase but will vanish in the trivial phase, i.e., we find
\begin{equation}
\left|\langle E_{-1}|\gamma_{1,a}|E_{1}\rangle\right|^{2}=\left|\langle E_{-1}|\sigma_{1}^{x}|E_{1}\rangle\right|^{2}\approx\begin{cases}
1 & \text{topological},\\
0 & \text{trivial}.
\end{cases}
\label{eq:MESO}
\end{equation}
The limiting factor for this approach is mainly the need to access both ground states $\ket{E_{\pm 1}}$, which comes at higher numerical costs than criterion 5 discussed below. Furthermore, when determining the ground states one has to ensure that one obtains the true ground states in the respective parity sectors and not some localized zero-energy state due to (strong) disorder. In addition, as exemplified in Fig.~\ref{fig:mu-dependence}(b) the Majorana edge state overlap possesses strong fluctuations inside the topological region which make a reliable extraction of the phase boundaries very hard.

4. A criterion often used to identify topological phases is the degeneracy of the entanglement spectrum\cite{Stoudenmire-11,Turner-11}, i.e., the spectrum of $\mathcal{H}=-\log \textrm{tr}_{L/2} \rho_{\textrm{gs}}$, with the ground-state density matrix $\rho_\text{gs}$, which is two-times degenerate in the topological phase. Hence its gap may be used as criterion, which turns out to be robust against disorder, i.e., it does not detect spurious localized zero-energy states. While this criterion is generally easily accessible with DMRG algorithms, it requires some effort to be numerically stable regarding the number of kept states $\chi$ (see Sec.~\ref{sub:Implementation} and Fig.~\ref{fig:chi}). To sample many disorder realizations in an affordable time we therefore did not apply this criterion. However, we show for an exemplary case in Fig.~\ref{fig:mu-dependence}(b) that it agrees with criterion 5 to be discussed in the following. 

5. Finally, let us devise the criterion we are going to use for our numerical calculations. As mentioned above, the topological phase corresponds to the magnetically ordered phase in the spin-chain representation \eqref{eq:Spin-Hamiltonian}, thus we start with the correlation function of the corresponding order parameter 
\begin{equation}
C_{r}^{i}=\sigma_{i}^{x}\sigma_{i+r}^{x}.
\label{eq:order}
\end{equation}
In the clean case, this function behaves as follows: In the trivial region $|\mu|>2t$ one finds that $C_{r}^{i}$ decays as\cite{YoungRieger96,RiegerIgloi97} $C_{r}^{i}=r^{-2x_{M}}\exp\left(-r/\xi\right)$ where $x_{M}\approx0.191$ and the correlation length is given by $\xi=1/\log(\mu/t)$. In contrast, in the topological region $|\mu|<2t$ the correlation function $C_{r}^{i}$ saturates at a constant value close to unity as $r\gg 1$. We will now use this behavior to construct a parameter which has a rather sharp transition at the critical point $\mu_{\textrm{c}}=2t$ and most importantly is robust against disorder.

To this end we consider the functions
\begin{equation}
G_{r}=\frac{1}{N-r}\sum_{i=1}^{N-r}\left|C_{r}^{i}\right|=\frac{1}{N-r}\sum_{i=1}^{N-r}\left|\sigma_{i}^{x}\sigma_{i+r}^{x}\right|,
\label{eq:our-parameter}
\end{equation}
i.e., we consider the mean of the correlations between the $N-r$ leftmost sites with the respective ones at distance $r$. For $r\gg1$ we obtain a sharp transition from $G_{r}\approx0$ in the trivial to $G_{r}\approx1$ in the topological phase. We note that both interactions [see Fig.~\ref{fig:mu-dependence}(a)] as well as spatial fluctuations in the individual samples due to the disorder [see Fig.~\ref{fig:mu-dependence}(b)] yield a reduction from the value $G_{r}\approx 1$, however, considering the sum of the correlation functions $C_r^i$ ensures that rare fluctuations on individual sites due to the disorder are smoothed out. 

While the correlation function \eqref{eq:order} is local in the spin-chain representation, the Jordan--Wigner string results in a non-local expression in terms of the fermionic degrees of freedom. For example, in terms of the Majorana operators one finds 
\begin{equation}
C_{r}^{i}=\left(-\ii\gamma_{i,b}\gamma_{i+r,a}\right)\prod _{j=i+1}^{i+r-1}\bigl(-\ii\gamma_{j,a}\gamma_{j,b}\bigr).
\end{equation}
For $r=1$ the Jordan--Wigner string obviously vanishes and $C_1^i$ just corresponds to the expectation value of finding Majorana fermions bound at neighboring sites. For $r\gg 1$, however, the string is essential to get a non-vanishing $C^i_r$ in the topological phase, reflecting the fact that the topological order is encoded in non-local properties of the fermionic model \eqref{eq:Kitaev-Hamiltonian}.

In Fig.~\ref{fig:mu-dependence} we compare the energy gap $\Delta E$, the gap in the entanglement spectrum $\Delta\mathcal{H}$, the Majorana edge state overlap \eqref{eq:MESO}, and the function $G_{r}$ for two examples of interacting systems. While for the clean system shown in Fig.~\ref{fig:mu-dependence}(a) all four criteria indicate the same critical chemical potential $\mu=\mu_\text{c}\approx 4t$, in the disordered system shown in Fig.~\ref{fig:mu-dependence}(b) the critical chemical potential extracted from the energy gap shows significant deviations due to localized zero-energy states. Similarly, the Majorana edge state overlap possesses strong fluctuations in the disordered system. On the other hand, the results obtained from the entanglement gap and $G_{r\gg 1}$ are identical, while the calculation of the latter turns out to be numerically less demanding (see Sec.~\ref{sub:Implementation} and Fig.~\ref{fig:chi}). We conclude that using the function \eqref{eq:our-parameter} provides a stable tool to detect the phase transition between the topological and trivial phases in the presence of both interactions and disorder. 

\section{Implementation}\label{sub:Implementation}
\begin{figure}[t]
\begin{center}
\includegraphics[width=0.9\columnwidth]{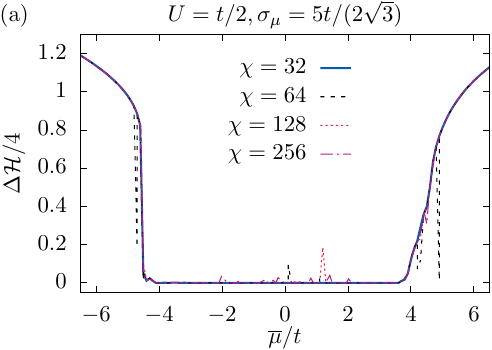}\vspace{5mm}
\includegraphics[width=0.9\columnwidth]{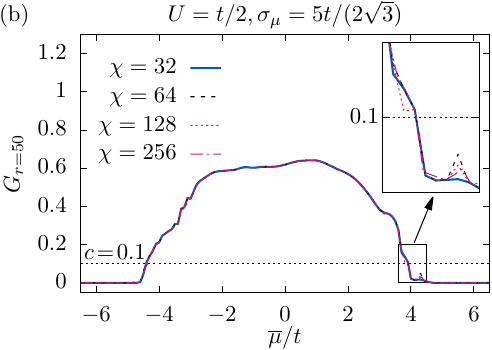}
\end{center}
\caption{(Color online) (a) Entanglement gap $\Delta\mathcal{H}$ and (b) function $G_{r=50}$ for an exemplary disorder configuration with mean value $\bar{\mu}$. The simulations were performed with $t=\Delta$ and $N=200$ for different values of $\chi$. We observe that for smaller matrix dimensions $\chi$ the entanglement gap may show accidental collapses, while the $\chi$-dependence of the function $G_{r=50}$ is much less significant.}
\label{fig:chi}
\end{figure}

As discussed in the previous section only the gap in the entanglement spectrum $\Delta\mathcal{H}$ and the function $G_{r\gg1}$ turned out to be robust with respect to strong disorder. The main advantage of the latter is that it is much less affected by truncation errors in the DMRG calculations, in particular in the strongly disordered case. This can be seen in Fig.~\ref{fig:chi}: The entanglement gap $\Delta\mathcal{H}$ shows fluctuations which may lead to accidental degeneracies in particular for small matrix dimensions $\chi$, thus increasing the uncertainty in the determination of the topological phase. In contrast, the result for the function $G_{r=50}$ only weakly changes with $\chi$, and yields reliable results\cite{footnote4} already for small matrix dimension $\chi=32$ thus reducing the time to simulate many disorder realizations. Hence in the following we use the criterion 
\begin{equation}
G_{r=50}(\bar{\mu})>c,\quad c=0.1,
\label{eq:criterion}
\end{equation}
to determine the topological phase. The function $G_{r=50}(\bar{\mu})$ depends implicitly on the mean $\bar{\mu}$ via the randomly distributed chemical potentials $\left\{ \mu_{i}\right\}$. The choice of $c$ is largely arbitrary, the effect of choosing other values will be discussed below. 

Practically, we proceed as follows:
\begin{enumerate}
\item We generate a set of randomly distributed chemical potentials $\left\{ \mu_{i}:i=1,\ldots,N\right\}$ from a uniform distribution with mean 0 and standard deviation $\sigma_{\mu}$. The random chemical potentials $\left\{ \mu_{i}\right\}$ are given as quenched disorder and are therefore not altered during the following procedure.
\item Now we introduce an auxiliary parameter $x$ by shifting the randomly distributed chemical potentials, i.e., $\mu_i\to x+\mu_i$, and determine the two points $\mu_{\text{c}}^\pm$ by evaluating the positions with $G_{r=50}(x=\mu_\text{c}^\pm)=c$. The critical mean value of the chemical potential is then set to $\mu_\text{c}=(\mu_\text{c}^+-\mu_\text{c}^-)/2$. The use of this symmetrized procedure nullifies random shifts of the center of the topological region, which have to average out to zero anyhow due to the symmetry of the problem under $\mu_i\to-\mu_i$. The function $G_{r=50}(x)$ is determined using a DMRG calculation with truncation at $\chi=32$ states.
\item We repeat step 1 for $M$ disorder realizations denoted by $\mu_i^j$, $j=1,\ldots,M$ (we use $M=100$ throughout). We use post-selection to ensure that the mean values $\nu^{j}=\sum_{i}\mu_{i}^{j}$ of the $M$ realizations represent the corresponding Gaussian distribution.\cite{footnote2} We only discard a drawn disorder configuration if already another one within the same acceptance limits for $\nu^{j}$ has been drawn before. Once we have obtained a valid disorder realization we calculate the critical chemical potential as in step 2, the final result for $\mu_\text{c}$ being the mean value of the critical chemical potentials of the $M$ disorder realizations.
\end{enumerate}

\begin{figure}[t]
\begin{centering}
\includegraphics[width=0.95\columnwidth]{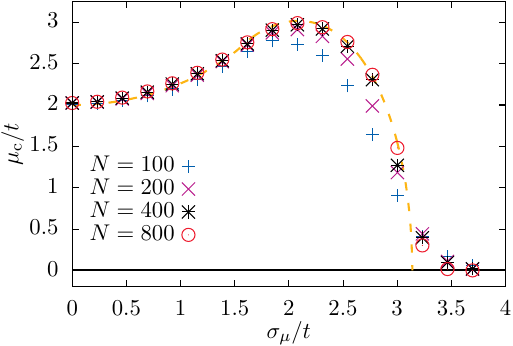}
\par\end{centering}
\caption{(Color online) Finite-size dependence of the critical chemical potential $\mu_\text{c}$ as a function of the disorder strength $\sigma_{\mu}$ for a non-interacting system with $t=\Delta$. The dashed line corresponds to the phase transition in the thermodynamic limit obtained from \eqref{eq:non-interacting-infinite}, which at weak disorder follows $\mu_\text{c}=2t+\sigma_\mu^2/(4t)+\mathcal{O}(\sigma_\mu^3)$. Numerically, $\mu_\text{c}$ has been determined with a matrix dimension of $\chi=32$ and $M=100$ disorder realizations using the algorithm described in Sec.~\ref{sub:Implementation}. The finite-size results approach the infinite-size behavior already for $N=400$. The vanishing of the topological phase for strong disorder strength is not as sharp as for the infinite-size case.}
\label{fig:disorder-clean}
\end{figure}

We decided to determine the critical chemical potential $\mu_\text{c}$ for each disorder realization, followed by averaging of $M$ realizations. Instead one may consider averaging the correlation functions over the disorder realizations, $\langle C_r^i\rangle_\text{avg}$, and then searching for the critical chemical potential. However, close to the critical point these averaged correlation functions are dominated by extremely rare\cite{RefaelFisher04} disorder realizations, which lead to a decay of $\langle C_{i}^{r}\rangle_{\textrm{avg}}\propto\exp\left(-3\pi^{2/3}r^{1/3}\right)$ in contrast to $\langle C_{i}^{r}\rangle_{\textrm{typ.}}\propto\exp\left(-\alpha r^{1/2}\right)$ for the typical correlation function [$\alpha$ is a disorder-dependent constant of order $\mathcal{O}\left(1\right)$]. Hence, one would have to take much more samples into account to obtain a decent computation of the average correlation functions $\langle C_r^i\rangle_\text{avg}$. Our comparison with exact results\cite{footnote3} in the thermodynamic limit for the non-interacting system (cf. Fig.~\ref{fig:disorder-clean}) confirmed that these extremely rare disorder realizations do not enter with an above-average weight in the determination of the disorder-averaged critical chemical potential $\mu_{\textrm{c}}$. 

\begin{table}[b]
\begin{center}
\begin{tabular}{|c|c|c|c|c|}
\hline
$U$ & $\sigma_\mu$ & $c=0.05$ & $c=0.1$ & $c=0.2$\\
\hline
0 & $2t/\sqrt{3}$ & $2.43\,t$ & $2.38\,t$ & $2.31\,t$\\
$t/2$ & $2t/\sqrt{3}$ & $4.35\,t$ & $4.27\,t$ & $4.13\,t$\\
0 & $4t/\sqrt{3}$ & $3.20\,t$ & $2.92\,t$ & $2.48\,t$\\
$t/2$ & $3.2\,t/\sqrt{3}$ & $3.61\,t$ & $3.17\,t$ & $2.51\,t$\\
\hline
\end{tabular}
\end{center}
\caption{Critical chemical potential $\mu_\text{c}$ for different values of the interaction $U$, disorder strength $\sigma_\mu$, and parameters $c$. The other parameters are $t=\Delta$, $N=400$, and $M=100$. For moderate disorder strengths $\sigma_\mu<t$ the errors in $\mu_\text{c}$ remain small.}
\label{tab:error}
\end{table}
We now briefly discuss the deviations in $\mu_\text{c}$ due to various sources of errors, namely approximations during the DMRG procedure as well as our choice of the parameter $c$ in \eqref{eq:criterion}. First, by varying the matrix dimension $\chi$ for various disorder realizations we found that a reasonable upper limit for the error $\delta\mu_{\text{c}}$ of $\delta\mu_{\text{c}}<0.04\,t$ can be achieved by using are rather small matrix dimension of $\chi=32$ [see Fig.~\ref{fig:chi}(b) for an example]. Second, we varied the value of $c$ and extracted the corresponding critical chemical potentials $\mu_\text{c}$ as shown in Tab.~\ref{tab:error}. As can be seen, the induced errors are comparable to the ones originating from the matrix dimension, at least for moderate disorder strengths $\sigma_\mu<t$.
\begin{figure}[b]
\begin{centering}
\includegraphics[width=0.95\columnwidth]{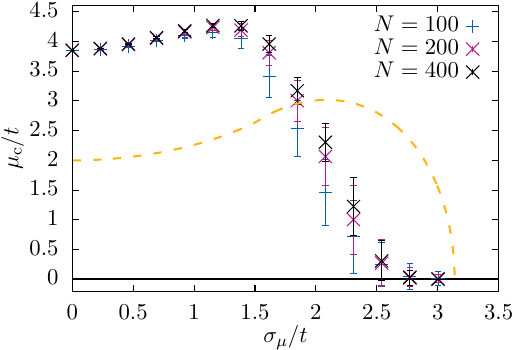}
\par\end{centering}
\caption{(Color online) Finite-size dependence of the critical chemical potential $\mu_\text{c}$ as a function of the disorder strength $\sigma_{\mu}$ for an interacting system with interaction strength $U=t/2$ and $t=\Delta$. The dashed line corresponds to the phase transition in the thermodynamic limit obtained from \eqref{eq:non-interacting-infinite}. $\mu_\text{c}$ has been determined using criterion \eqref{eq:criterion} with a matrix dimension of $\chi=32$ and $M=100$ disorder realizations. The error bars show the statistical error defined as half of the standard deviations.}
\label{fig:disorder-interaction}
\end{figure}

On top of these systematical errors each critical chemical potential will have a statistical error stemming from the fact that by choosing $N$ random variables for a given quenched finite-size disorder one will deviate from the smooth distribution obtained in the thermodynamic limit. We plot the error bars due to this statistical error, which we define as half of the respective standard deviations, in Figs.~\ref{fig:disorder-interaction}--\ref{fig:pd2}. We note that the statistical error only vanishes when taking the thermodynamic limit, $N\to\infty$, followed by the limit of infinite number of disorder realizations, $M\to\infty$. 

In addition to the statistical error at finite $N$ and $M$, one has the usual finite-size error on the critical chemical potential. This is illustrated in Fig.~\ref{fig:disorder-clean} for the clean system. We observe, however, that even though we used a small matrix dimension $\chi=32$ to reduce the numerical workload for averaging over the disorder samples, a remarkably good agreement between the finite-size results for larger chains ($N=400$ and $N=800$) and the exact result in the thermodynamic limit is obtained, even in the strongly disordered regime. Generally we observe that considering smaller systems results in a reduction of the topological phase, even in the interacting case as shown in Fig.~\ref{fig:disorder-interaction}. 

\section{Interplay of disorder and interaction}\label{sec:Interplay-of-disorder}

\begin{figure}[t]
\begin{centering}
\includegraphics[width=0.95\columnwidth]{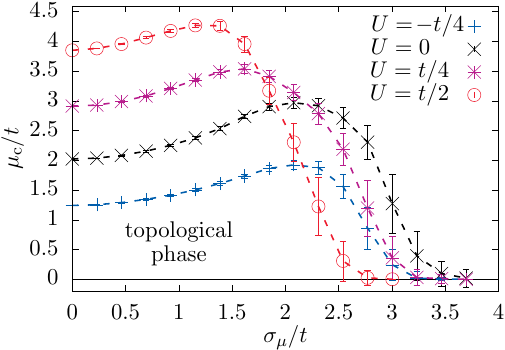}
\par\end{centering}
\caption{(Color online) Critical chemical potential $\mu_\text{c}$ as a function of the disorder strength $\sigma_{\mu}$ for different interaction strengths and $t=\Delta$. We used $N=400$, $M=100$, and $\chi=32$. The dashed lines connecting the data points are a guide to the eye. We observe that repulsive interactions stabilize the topological phase for moderate disorder $\sigma_\mu<t$, while at strong disorder the interactions suppress the topological phase. In contrast, attractive interactions destabilize the topological phase for all disorder strengths. The error bars show half of the standard deviation.}
\label{fig:disorder-interaction-compare}
\end{figure}

\begin{figure}[t]
\begin{centering}
\includegraphics[width=0.95\columnwidth]{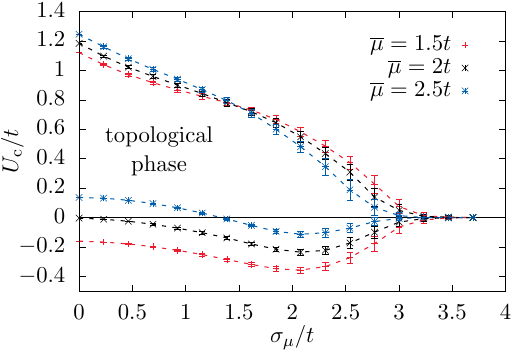}
\par\end{centering}
\caption{(Color online) Critical interaction strength $U_\text{c}$ as a function of the disorder strength $\sigma_{\mu}$ for different mean values of the chemical potential $\bar{\mu}$ and $t=\Delta$. We used $N=400$, $M=100$, and $\chi=32$. The dashed lines connecting the data points are a guide to the eye. Again we observe that weak interactions and weak to moderate disorder strengths stabilize the topological phase. The error bars show the statistical error, which is much smaller than in Fig.~\ref{fig:disorder-interaction-compare}.}
\label{fig:pd2}
\end{figure}

We now turn to the main result of our work, namely the determination of the phase diagram in the presence of disorder and interactions. Before doing so we point out that the phase diagrams presented in Figs.~\ref{fig:disorder-interaction-compare} and~\ref{fig:pd2} have to be understood as averaged ones for finite systems in the following sense: While for each disorder realization we have a fixed Hamiltonian and thus a definitive answer whether Majorana modes are present or not, the shown phase diagrams are obtained by averaging over $M$ disorder realizations. Thus in the region denoted ``topological phase" a finite percentage of disorder realizations will possess Majorana modes. The precise value of this percentage depends on the specific point in the phase diagram and the system size, while its value at the indicated phase boundaries also depends on the details of our approach, like for example the value of $c$ chosen in the criterion \eqref{eq:criterion}. However, we expect that in the thermodynamic limit the indicated phase boundaries become sharp transitions.

Now we turn to the discussion of the obtained phase diagrams. The finite-size behavior for the interplay of disorder and interactions is illustrated in Fig. \ref{fig:disorder-interaction}, where the results for the critical chemical potential $\mu_{c}$ for chains with up to $N=400$ sites are shown. Generally, the finite-size effects are very similar to the non-interacting case. For example, at weak to moderate disorder, $\sigma_\mu<t$, the finite-size effects are very weak. We also observe that the statistical errors decrease quickly when increasing the system size, and are only appreciable in the strongly  disordered regime.

We have performed simulations for systems with $t=\Delta$ and several values of the interaction strength for chains of length $N=400$ using $\chi=32$ states and $M=100$ disorder realizations. Our results for the phase diagram are summarized in Figs.~\ref{fig:disorder-interaction-compare} and~\ref{fig:pd2}, which constitute the main result of our work. 

For weak disorder strengths $\sigma_{\mu}/t\lesssim1$ we observe in Fig.~\ref{fig:disorder-interaction-compare} that the topological phase is stabilized by the disorder irrespective of the interaction strength, i.e., for all considered values of $U$ the critical chemical potential $\mu_\text{c}$ initially increases with $\sigma_\mu$. In fact, in this regime disorder and interactions seem not to influence each other and the phase boundary is well described by simply adding up their individual effects. Thus for $t=\Delta$ and in the thermodynamic limit we obtain
\begin{equation}
\mu_\text{c}=2t+\frac{32U}{3\pi}+\frac{\sigma_\mu^2}{4t}+\mathcal{O}(U^2,\sigma_\mu^3),
\label{eq:mc}
\end{equation}
which is consistent with the numerically determined phase boundary shown in Fig.~\ref{fig:disorder-interaction-compare}. 
\begin{figure}[t]
\begin{centering}
\includegraphics[width=0.95\columnwidth]{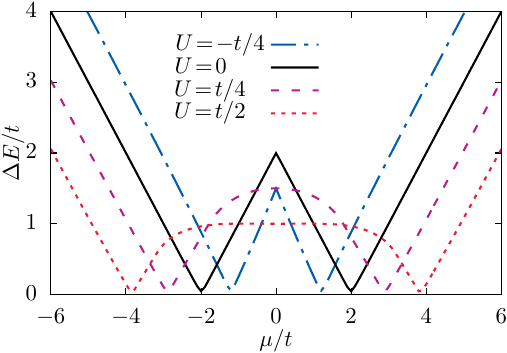} 
\par\end{centering}
\caption{(Color online) Energy gap between the ground state and the second excited state for non-disordered systems with chemical potential $\mu$ and $t=\Delta$, obtained from DMRG simulations with $\chi=32$ and $N=400$. We observe that the energy gap in the topological phase (centered around $\mu=0$) is reduced by both repulsive and attractive interactions.} 
\label{fig:gap-mu-dependence}
\end{figure}

Similarly, in the phase diagram shown in Fig.~\ref{fig:pd2} we observe that for weak interactions moderate disorder stabilizes the topological phase. In fact, the phase boundary is just described by \eqref{eq:mc}. As discussed in Sec.~\ref{sub:Interacting-case-without-disorder}, for strong interactions the clean system possesses incommensurate and commensurate charge-density wave phases with the transition from the topological phase to the incommensurate charge-density phase at $U_\text{c}\approx t$ (for $\mu\approx 2t$). Here we see that even weak disorder suppresses the topological phase by decreasing the critical interaction strength $U_\text{c}$ as compared to the non-interacting system, in contrast to the transition at weak interaction strengths $U\approx 0$.

Stronger disorder generally leads to a collapse of the topological phase since the system is presumably dominated by states localized around sites with strongly negative $\mu_{i}$. Interactions amplify this behavior, i.e., both for repulsive and attractive interactions the topological phase is destroyed already at weaker disorder. This is consistent with the findings for disordered, superconducting nanowires with repulsive interactions.\cite{Lobos-12,Crepin-14} An intuitive explanation for the suppressive effect of interactions can be given by considering the energy gap above the two-fold degenerate ground state in the topological phase. As illustrated in Fig.~\ref{fig:gap-mu-dependence} both repulsive and attractive interactions decrease this gap, which allows sufficiently strong fluctuations in the chemical potential to destroy the topological phase more easily.

\section{Conclusion}
We have studied the combined effect of disorder and interactions on the phase diagram of a generalized Kitaev/Majorana chain. To this end we considered several observables for the determination of the phase boundary between the topological phase and the trivial phase. It turned out that the most useful observable was given by the function \eqref{eq:our-parameter}, which was calculable in DMRG simulations with relatively small numerical effort. We observed that moderate disorder stabilizes the topological order even in the model with weak to moderate interactions. However, the suppression of the topological phase by strong disorder was found to be amplified by both repulsive and attractive interactions as can be seen in Fig.~\ref{fig:disorder-interaction-compare}.

\acknowledgments 
We thank Philippe Corboz, Fabian Hassler, Fabian Heidrich-Meisner, Christoph Karrasch, Hosho Katsura, Dante Kennes, Salvatore Manmana, and Tatjana Pu\v{s}karov for useful comments and discussions. This work is part of the D-ITP consortium, a program of the Netherlands Organisation for Scientific Research (NWO) that is funded by the Dutch Ministry of Education, Culture and Science (OCW).


\begin{thebibliography}{10}
\bibitem{footnote4}
We expect that the calculation of other quantities (e.g. the precise determination of the decay of correlation functions) will require matrix dimensions much larger than $\chi=32$, especially at the critical point.
\bibitem{footnote2}
This is obtained as approximation from the Irwin--Hall distribution via the central limit theorem.
\bibitem{footnote3}
For exemplary disorder realisations in the non-interacting system we have explicitly checked that the function $G_{r}$ calculated using DMRG agrees with the exact result obtained from the solution\cite{Pfeuty70} of the Ising chain.
\end{thebibliography}

\end{document}